\documentclass[nofootinbib,preprint,superscriptaddress]{revtex4-1}
\usepackage{amsmath,graphicx,hyperref}

\newcommand{\fms}[1]{{#1}\!\!\!/}

\newcommand{\mc}{\mathcal}
\newcommand{\mr}{\mathrm}

\newcommand{\be}{\begin{equation}} 
\newcommand{\ee}{\end{equation}} 
\newcommand{\bea}{\begin{eqnarray}} 
\newcommand{\eea}{\end{eqnarray}}

\newcommand{\dg}{\dagger}
\newcommand{\n}{\overline{n}}
\newcommand{\nn}{\frac{\fms{\overline{n}}}{2}} 
\newcommand{\nnn}{\frac{\fms{n}}{2}}

\newcommand{\blp}[1]{{\bf{#1}}_{\perp}}

\newcommand{\nnb}{\nonumber} 

\newcommand{\as}{\alpha_s}

\begin{document}

%\vspace*{18pt}

%%%%%%%%%%%%%%%%%%%%%%%%%%%%%%%%%%%%%%%%%%%%%%%%%%%%%%%%%%%%%%%%%%%%%%
%%%%%%%%%%%%%%%%%%%%%%%%%%%%% Title %%%%%%%%%%%%%%%%%%%%%%%%%%%%%%%%%%
%%%%%%%%%%%%%%%%%%%%%%%%%%%%%%%%%%%%%%%%%%%%%%%%%%%%%%%%%%%%%%%%%%%%%%

\title{Exclusive heavy quark dijet cross section}

\def\Seoultech{Institute of Convergence Fundamental Studies and School of Liberal Arts, Seoul National University of Science and Technology, Seoul 01811, Korea}

\author{Chul Kim}
\email[E-mail:]{chul@seoultech.ac.kr}
\affiliation{\Seoultech\vspace{0.7cm}}

\begin{abstract} \vspace{0.1cm}\baselineskip 3.0ex 
We study the exclusive heavy quark dijet cross section from $e^+e^-$-annihilation using soft-collinear effective theory.  
In order to resum the large logarithms of small jet veto parameter $\beta$ and jet radius $R$, we factorize the cross section into the hard, hard-soft, collinear, and collinear-soft parts. Compared with the case of a massless quark, the jet sector with the collinear and collinear-soft parts can be modified to include the heavy quark mass. The factorization of the jet sector can be systematically achieved through matching onto the boosted heavy quark effective theory. 
Heavy quark mass corrections enhance the cross section sizably and cannot be ignored when the quark mass is comparable with the jet size $E_JR$. We also analyze the exclusive heavy quark pair production in the limit as $R$ goes to zero. 
Using the resummed result, the top isolation effects on the cross section are estimated.

\end{abstract}

\maketitle 

%%%%%%%%%%%%%%%%%%%%%%%%%%%%%%%%%%%%%%%%%%%%%%%%%%%%%%%%%%%%%%%%%%%%%%
\baselineskip 3.5ex 

\section{Introduction}

Jets, collimated beams of strongly interacting particles, are an important observable for scrutinizing the Standard Model and for finding new physics signals. Jets are usually characterized by the jet energy ($E_J$) and radius ($R$). If $R$ is small, the jets can be handled independently of the hard interactions. Hence QCD factorization to separate short and long distance physics becomes an important tool to study the jet physics, and it enables us to systematically resum large logarithms of small $R$ that appears in the scattering cross section with the jet~\cite{Dasgupta:2014yra,Kang:2016mcy,Dai:2016hzf}.   

If a jet with small radius $R$ contains a heavy quark, the heavy quark mass could be comparable with 
a jet size, which is roughly given by $E_JR$. 
In this case the finite size of the heavy quark mass can give significant corrections to the predictions of jets in which the heavy quark has been taken as a massless parton. Therefore understanding the heavy quark mass effects is an important ingredient 
for a precise estimation of the jet, and furthermore for probing electroweak and new physics since the heavy quark is sensitive to Yukawa couple.

In this paper we study the exclusive heavy quark dijet scattering cross section in $e^+e^-$-annihilation. 
Basically the dijet cross section does not become additionally singular when we take the massless limit on the heavy quark. 
Therefore the dijet cross section can be a good testing system for investigating heavy quark mass effects by comparing both the massive and massless cases. 
The dijet cross section depends on the jet veto parameter $\beta$ and the radius $R$. 
Since both the parameters are small and produce large logarithms, the perturbative expansion at the fixed order in $\as$ is not reliable. Hence 
we consider the resummation of the cross section employing soft-collinear effective theory (SCET)~\cite{Bauer:2000ew,Bauer:2000yr,Bauer:2001yt,Bauer:2002nz}.  

The organization of the paper is as follows: In section \ref{sec2} we discuss the factorization of the heavy quark dijet cross section. In section \ref{sec3}, using the factorization theorem we resum large logarithms and estimated the heavy quark mass impact on the cross section. In section \ref{sec4}, taking the limit of the cross section as $R \to 0$, we consider the exclusive  
heavy quark pair production. Finally we conclude in section \ref{sec5}.

\section{Factorization of the dijet cross section} 
\label{sec2}

For construction of dijet events in $e^+e^-$-annihialation, we apply the Sterman-Weinberg (SW) algorithm~\cite{Sterman:1977wj}. In the SW algorithm, energetic particles that are deposited within a cone with the half angle $R/2$ constitute a jet. 
At next-to-leading order (NLO) in $\as$, if the angle $\theta$ between two energetic particles satisfy 
\be
\theta < R,  
\ee
they merge to a jet. So this constraint is the same as the ones for inclusive $\mr{k_T}$-type algorithms~\cite{Catani:1993hr,Ellis:1993tq,Dokshitzer:1997in,Cacciari:2008gp}.  
In addition, to be infrared (IR) safe, soft particles with energies less than $\beta Q$ are included in the dijet events. Here $Q$ is the center of mass energy of the incoming electron and positron, and the veto parameter $\beta$ is given to be small, i.e, $\beta \ll 1$.  

With small $\beta$ and $R$ adopted, the dijet cross section with massless partons has been studied in the framework of SCET, and its factorization theorem is formulated as~\cite{Cheung:2009sg,Ellis:2010rwa,Chay:2015ila}
\be
\label{dijetf0} 
\sigma_2 (Q,\beta, R) = \sigma_0 H(Q,\mu) \mc{J}_n (E_JR,\mu) \mc{J}_{\n} (E_JR,\mu) S (\beta Q, R,\mu),
\ee
where $\sigma_0$ is the Born level cross section, and the jet energy $E_J$ can be given by $Q/2$. $H$ is the hard function, $\mc{J}_{n(\n)}$ is the integrated jet function to describe $n(\n)$-collinear interactions inside the jet, and $S$ is the soft function for soft gluon radiations that depends on the jet veto. 
To NLO order in $\as$, each factorized function in Eq.~\eqref{dijetf0} is given by 
\bea 
\label{Hnlo}
H(Q,\mu) &=& 1 + \frac{\as C_F}{2\pi} \Bigl(-3\ln \frac{\mu^2}{Q^2} - \ln^2 \frac{\mu^2}{Q^2} - 8 + \frac{7\pi^2}{6} \Bigr), \\
\label{IJ0nlo}
\mc{J}_n (E_JR,\mu) = \mc{J}_{\n} (E_JR,\mu) &=& 1+\frac{\as C_F}{2\pi} \Bigl(\frac{3}{2} \ln \frac{\mu^2}{E_J^2R^2} + \frac{1}{2} \ln^2 \frac{\mu^2}{E_J^2R^2} + \frac{13}{2} - \frac{\pi^2}{4}\Bigr), \\
\label{nSnlo}
S (\beta Q,R,\mu) &=& 1+\frac{\as C_F}{2\pi} \Bigl(4\ln \frac{\mu^2}{4\beta^2Q^2} \ln \frac{R}{2} -4\ln^2\frac{R}{2} -\frac{\pi^2}{3}\Bigr).
\eea

All the large logarithms in $H$ and $\mc{J}_{n,\n}$ are minimized by the renormalization scale choices of $\mu_h \sim Q$ for $H$ and $\mu_{c} \sim E_JR$ for $\mc{J}_{n,\n}$. However, the large logarithms of $R$ in the soft function persist even if we set the soft scale as $\mu_{s} \sim 2\beta Q$. Therefore the renormalization group (RG) evolution for $S$ from the factorization scale $\mu$ to the soft scale $\mu_s$ does not completely resum all the possible large logarithms, hence we need to additionally factorize $S$ to capture scales to minimize all the logarithms. 
For this, we can subdivide soft interactions into the `hard-soft (hsoft)' and the `collinear-soft (csoft)' interactions. The corresponding modes of gluons scale as 
\bea 
\label{hsm}
p_{hs} &=& (p_{hs}^+, p_{hs}^{\perp}, p_{hs}^-) \sim Q\beta (1,1,1),  \\
\label{csm}
p_{n,cs} &\sim& Q\beta (1,R,R^2), \  \ p_{\n,cs} \sim Q\beta (R^2,R,1),
\eea
where $p_+ \equiv \n\cdot p$ and $p_- \equiv n\cdot p$. The two lightcone vectors, $n$ and $\n$, are back-to-back and satisfy $n\cdot \n =2$. The hsoft mode for Eq.~\eqref{hsm} is responsible for wide angle soft radiations, hence cannot resolve the jet boundary with the radius $R$. However, two csoft modes in Eq.~\eqref{csm} radiate over narrow angles around both the jet axes and can recognize the jet boundary.  

The refactorization of soft interactions can be performed similarly to the conventional factorization into hard and collinear interactions from full theory. At scale $\mu \sim \beta Q$, we first integrate out the hsoft mode matching onto the lower effective theory with the csoft modes, and obtain the hsoft function. Then at the lower scale, $\mu \sim \beta Q R \ll \beta Q$, the remaining two csoft modes cannot communicate each other, and thus factorizes. 

As a result, the soft function $\mc{S}$ in Eq.~\eqref{nSnlo} can be factorized into the hsoft, $n$- and $\n$-csoft functions such as~\cite{Becher:2015hka,Chien:2015cka} 
\be
S(\beta Q, R,\mu) = S_{hs} (2\beta Q,\mu) \mc{S}_{n} (\beta Q R,\mu) \mc{S}_{\n} (\beta Q R,\mu).
\ee
Here the NLO results for the factorized functions are given by 
\bea
\label{Shs1} 
S_{hs} (2\beta Q, \mu) &=& 1+ \frac{\as C_F}{2\pi} \Bigl(\ln^2 \frac{\mu^2}{4\beta^2Q^2}-\frac{\pi^2}{2}\Bigr), \\
\label{Sn1}
\mc{S}_n (\beta Q R,\mu) = \mc{S}_{\n} (\beta Q R,\mu) &=& 1 - \frac{\as C_F}{2\pi} \Bigl( \frac{1}{2} \ln^2\frac{\mu^2}{\beta^2 Q^2 R^2}+\frac{\pi^2}{12}\Bigr). 
\eea 
Therefore the complete factorization theorem for the dijet cross section is given as 
\bea
\sigma_2 (Q,\beta, R) &=& \sigma_0 H(Q,\mu) S_{hs}(2\beta Q,\mu) \nnb \\
\label{dijetf1} 
&&\times \Bigl[\mc{J}_n (E_JR,\mu) \mc{S}_n(2\beta E_J R,\mu) \Bigr]
\Bigl[\mc{J}_{\n} (E_JR,\mu) \mc{S}_{\n} (2\beta E_J R,\mu) \Bigr].
\eea

The factorization theorem, Eq.~\eqref{dijetf1}, can be also applied to the heavy quark dijet cross section that is based on heavy quark pair production. To do so, the jet sector, $\mc{J}_{n(\n)}\mc{S}_{n(\n)}$, needs to be modified to include the heavy quark mass. The produced energetic heavy quarks leading to jets have collinear interactions basically, and the momenta of the heavy quarks in $n$ and $\n$ directions scale as 
\be
\label{spc}
p_n = (p_n^+,p_n^{\perp},p_n^-) \sim E_J(1,R,R^2),\  \ p_{\n} \sim E_J(R^2,R,1).
\ee
We will consider the heavy quark mass $m$ in the limit, $m \lesssim E_JR \ll E_J$, so the offshellnesses of the heavy quarks scale as $p_n^2 \sim p_{\n}^2 \sim E_J^2 R^2 \gtrsim m^2$. 

These collinear interactions of the heavy quark can be described by the massive version of SCET, i.e., $\mr{SCET_M}$~\cite{Leibovich:2003jd,Rothstein:2003wh,Chay:2005ck}. However, the jet veto dependences on $\beta$ are not effectively resolved by purely collinear interactions, hence we need the csoft modes to capture the veto dependences.  
The scaling behavior of the csoft modes have been described in Eq.~\eqref{csm}. Hence we notice that these csoft modes can be  also subsets of the collinear modes in Eq.~\eqref{spc}. 

When we separate the csoft interactions from the collinear interactions in the heavy quark sector, we can introduce the boosted heavy quark effective theory (bHQET), i.e, the boosted version of HQET. For example, let us consider an energetic heavy quark moving in $n$ direction. With collinear interactions integrated out, at the lower scale $\mu \sim Q\beta R$, the heavy quark  only has csoft interactions. The heavy quark momentum can be written as 
\be 
p^{\mu} = mv^{\mu} + k^{\mu},  
\ee 
where $v^{\mu}$ is the heavy quark velocity to be normalized as $v^2=1$, and $k^{\mu}$ is a residual csoft momentum. Under the csoft interactions, the velocity does not change. Since $mv^{\mu}$ is $n$-collinear momentum, the velocity scales as $v^{\mu} = (v_+,v_{\perp},v_-) \sim (1/\lambda,1,\lambda)$, where $\lambda \sim m/p_+$. Conveniently, if we choose the frame for $v_{\perp}$ to be zero, the velocity $v$ can be given by
\be
\label{velo} 
v^{\mu} = v_+ \frac{n^{\mu}}{2} + v_- \frac{\n^{\mu}}{2} = v_+ \frac{n^{\mu}}{2} + \frac{1}{v_+} \frac{\n^{\mu}}{2}. 
\ee 

To construct bHQET from $\mr{SCET_M}$, we first integrate out collinear interactions, i.e., collinear gluons, then match the heavy quark collinear field $\xi_n$ in $\mr{SCET_M}$ onto the bHQET field,  
\be
\label{Qmatch}
\xi_n (x) = \sqrt{\frac{v_+}{2}} e^{-imv\cdot x} h_n (x). 
\ee
Thus, the bHQET field $h_n$ has the same spinor property as $\xi_n$ and satisfies 
\be 
\fms{n} h_n =0,~~\frac{\fms{n}\fms{\n}}{4} h_n = h_n.
\ee
This preserves the power counting with respect to large energy that has been applied to $\mr{SCET_M}$.      
As a result, bHQET at leading power in $1/m$ is 
\be
\label{LObHQET}
\mc{L}_{\mr{bHQET}}^{(0)} = \bar{h}_n v\cdot iD_{cs} \nn h_n.
\ee
For more details of bHQET Lagrangian, we refer to Ref.~\cite{DKL}. 

Therefore the factorization of the heavy quark jet sector can be performed through matching onto bHQET, and the result for the $n$-collinear jet is expressed as  
\be 
\label{facthqj}
\mc{J}_{Q,n} (E_JR,m,\mu) \mc{S}_{Q,n} (2\beta, E_JR,m,\mu), 
\ee 
where the subscript `$Q$' denotes the heavy quark. The $\n$-collinear heavy quark jet sector in the opposite direction can be factorized in the same way.  

In Eq.~\eqref{facthqj}, $\mc{J}_{Q,n}$ is the integrated heavy quark jet function (iHQJF)~\cite{Dai:2018ywt}, which is the result of integrating out collinear gluon radiations inside the jet. 
At NLO in $\as$, $\mc{J}_{Q,n}$ is given by~\cite{Dai:2018ywt}
\bea
\mc{J}_{Q,n} (E_JR,m,\mu) &=& \mc{J}_{Q,\n} (E_JR,m,\mu) =1+ \frac{\as C_F}{2\pi} \Biggl[ \frac{3+b}{2(1+b)} \ln\frac{\mu^2}{B^2} +\frac{1}{2} \ln^2 \frac{\mu^2}{B^2}+ f(b)+g(b) \nnb \\
\label{IJFnlo}
&&+ \frac{1}{1+b}\bigl(2+\ln(1+b)\bigr) 
-\frac{1}{2} \ln^2 (1+b) -\mr{Li}_2 (-b) + 2 -\frac{\pi^2}{12} \Biggr],
\eea
where $b \equiv m^2/(E_J^2R^2)$ and $B=\sqrt{E_J^2R^2+m^2}$. The functions $f(b)$ and $g(b)$ have integration forms, 
\bea
\label{fb}
f(b) &=& \int^1_0 dz \frac{1+z^2}{1-z} \ln\frac{z^2+b}{1+b}, \\
\label{gb}
g(b) &=& \int^1_0 dz \frac{2z}{1-z}\Bigl(\frac{1}{1+b}-\frac{z^2}{z^2+b}\Bigr). 
\eea
In the limit $b\to 0~(m\to 0)$, these functions are $f(0) = 5/2-2\pi^2-3$ and $g(0) = 0$. In the limit $b$ goes to an infinity, corresponding to $R\to 0$, they go to $f(\infty) = g(\infty) =0$. 

The heavy quark csoft function $\mc{S}_{Q,n}$ in Eq.~\eqref{facthqj} is analyzed in bHQET and is defined as 
\be
\label{hqcsoft}
\mc{S}_{Q,n} (2\beta,E_JR,m,\mu) = \frac{1}{2N_c}  \sum_s \sum_{X_{cs} \in \sigma_2} \mr{Tr} \frac{v_+}{2p_J^+} \langle 0 |~ Y_{\n}^{cs\dg} h_n ~|Q_s X_{cs}\rangle 
\langle Q_s X_{cs}|~ \bar{h}_n Y_{\n}^{cs} \nn ~| 0 \rangle,
\ee
where $v_+ = p_J^+/m\sim 2E_J/m$, $Q_s$ is the heavy quark with spin $s$, and $X_{cs}$ is the csoft final state, which should be in the phase space that the heavy quark jet and the veto cover. $Y_{\n}^{cs}$ is the csoft Wilson line, where $n$-csoft gluon radiations from other sectors have been eikonalized as  
\be
Y_{\n}^{cs}(x)=\mr{P} \exp\Bigl[ig\int^{\infty}_{x} ds~ \n\cdot A_{n,cs} (sn)\Bigr].
\ee
Here `P' denotes the path ordering, and $A_{n,cs}^{\mu}$ is the csoft gluon propagating in $n$ direction. 
From Eq.~\eqref{Qmatch}, the spin sum rule for the bHQET field is given by 
\be
\label{spinsum} 
\sum_s h_n |Q_s (p_+) \rangle  \langle Q_s (p_+) | \bar{h}_n = 2m \nnn = m \fms{n}. 
\ee
So the csoft function at tree level is normalized as $\mc{S}_{Q,n}^{(0)} = 1$. 

\begin{figure}[h]
\begin{center}
\includegraphics[height=6cm]{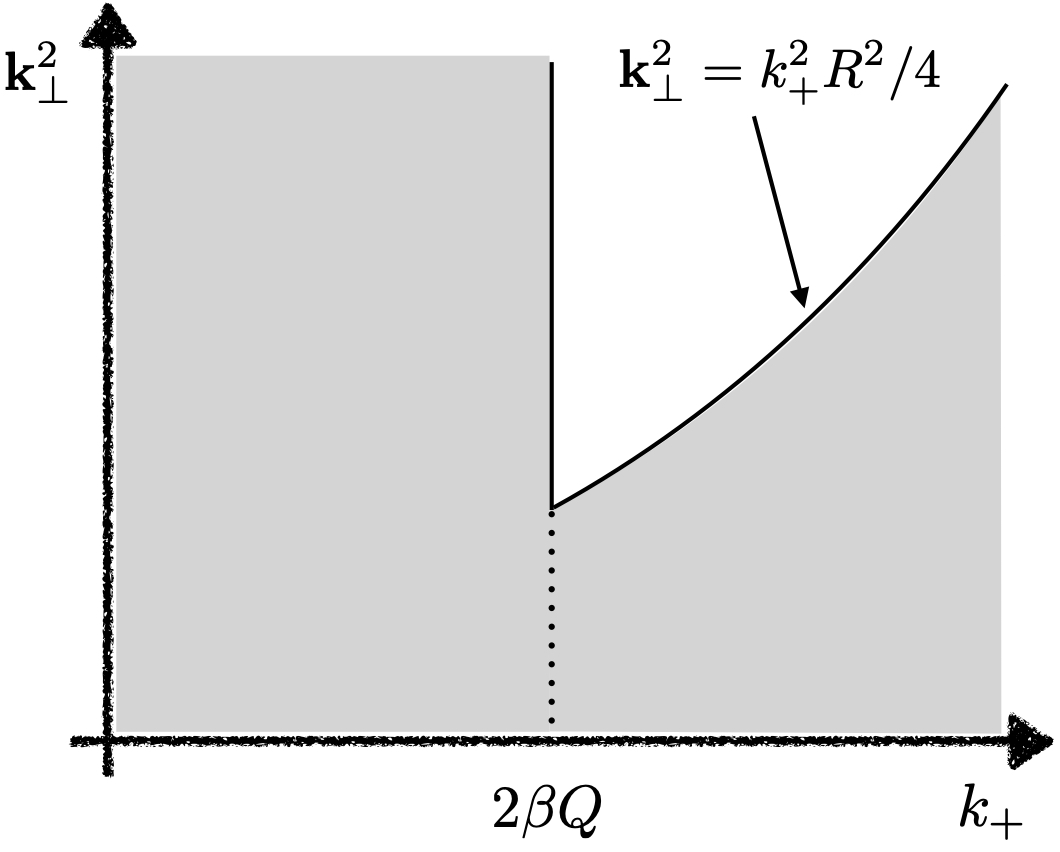}
\end{center}
\vspace{-0.8cm}
\caption{\label{fig1}\baselineskip 3.5ex
The available phase space for real radiations of $n$-csoft gluon in the dijet cross section at one loop.   }
\end{figure}

At one loop order, the available phase space for a radiated csoft gluon is illustrated in Fig.~\ref{fig1}. 
The momentum constraint for the gluon to be inside a jet is given by
\be 
\blp{k}^2 < \frac{R^2}{4}k_+^2. 
\ee
And, due to the jet veto constraint, the gluon satisfying the condition, $k_+ \sim 2k^0 < 2\beta Q$, can be counted as part of the dijet events even if it is outside the jet. 

In Fig.~\ref{fig1}, on the $\blp{k}^2=0$ axis we obtain the logarithm of $m$, which becomes singular as $m\to 0$. 
The soft IR divergence arises on the $k_+=0$ axis. Note that overall one loop result including virtual contributions is given by zero if the real gluon emission covers the full phase space without the dijet event constraint. 
This indicates that the virtual contributions can be considered as the negative contribution of the real radiations with the full phase space. So, when the virtual contributions are combined, nonvanishing contributions come from the outside of the shaded region in Fig.~\ref{fig1}. Hence the net result has only ultraviolet (UV) divergences without the IR divergence or the term with $\ln m$. 

As a result, we obtain the csoft function $\mc{S}_{Q,n}$ to NLO in $\as$ as 
\bea 
\mc{S}_{Q,n} (2\beta, E_JR,m,\mu) &=& 1 + \frac{\as C_F}{2\pi} \Bigl[-\frac{1}{2} \ln^2\frac{\mu^2}{4\beta^2B^2} + \frac{b}{1+b} \ln\frac{\mu^2}{4\beta^2B^2} - \frac{\ln(1+b)}{1+b} \nnb \\
\label{csoftnlo}
&& \ \ \ \ \ \ \ \ \ 
+\frac{1}{2} \ln^2(1+b) + \mr{Li}_2 (-b) + \frac{\pi^2}{12} \Bigr]. 
\eea 
This is a new result from this paper. $\mc{S}_{Q,\n}$ for $\n$-csoft interactions has the same result. If we take the limit $m\to 0$, Eq.~\eqref{csoftnlo} becomes the massless result shown in Eq.~\eqref{Sn1}. 

Finally, we confirm that the heavy quark dijet cross section can be factorized as 
\bea
\sigma_2 (Q,\beta, R,m) &=& \sigma_0 H(Q,\mu) S_{hs}(2\beta Q,\mu) 
\Bigl[\mc{J}_{Q,n} (E_JR,m,\mu) \mc{S}_{Q,n}(2\beta, E_J R,m,\mu) \Bigr]
\nnb \\
\label{hqdijetf} 
&&\times 
\Bigl[\mc{J}_{Q,\n} (E_JR,m,\mu) \mc{S}_{\n} (2\beta, E_J R,m,\mu) \Bigr],
\eea
where $E_J = Q/2$. We might be able to consider the heavy quark mass correction to the hard function $H$, but it can be safely ignored since it is suppressed by $m^2/Q^2$. Compared with the massless case, the hsoft function also remains unchanged since hsoft radiations are insensitive to the quark mass. If we consider the limit $m\to 0$ in Eq.~\eqref{hqdijetf}, the result recovers Eq.~\eqref{dijetf1}. This is a good consistency check for the heavy quark cross section. Furthermore we can apply Eq.~\eqref{hqdijetf} to the limit $E_JR \gg m$. In this case the result can be considered as the one with all power corrections in the expansion of $(m^2/E_J^2R^2)^n$. 

\section{Resummation of large logarithms in the heavy quark dijet cross section}
\label{sec3}

In Eq.~\eqref{hqdijetf} each factorized function has its own scale to minimize large logarithms. So, through RG evolution of the factorized functions from the factorization scale to their own scales, we can consistently resum the large logarithms. The factorized functions satisfy the following RG equations, 
\be 
\label{RGE}
\frac{df}{d\ln\mu} = \gamma_f f,~~~f=H,S_{hs},\mc{J}_{Q}, \mc{S}_{Q}. 
\ee
Here $\mc{J}_Q \equiv \mc{J}_{Q,n (\n)}$, and $\mc{S}_Q \equiv \mc{S}_{Q,n (\n)}$. 
To next-to-leading logarithm (NLL) accuracy needed to resum contributions of order unity, the anomalous dimensions 
are given by  
\bea 
\label{g1}
\gamma_h &=& -2 \Gamma_C (\as) \ln\frac{\mu^2}{Q^2} - \frac{3\as C_F}{\pi},~~\gamma_{hs} = 2\Gamma_C (\as) \ln\frac{\mu^2}{4\beta^2Q^2}, \\
\label{g2}
\gamma_c &=& \Gamma_C (\as) \ln\frac{\mu^2}{B^2} + \frac{\as C_F}{2\pi}\frac{3+b}{1+b},~~ \gamma_{cs} = \Gamma_C (\as) \ln\frac{\mu^2}{B^2} + \frac{\as C_F}{\pi}\frac{b}{1+b},  
\eea
where $\gamma_{c}$ is for $\mc{J}_{Q}$ and $\gamma_{cs}$ is for $\mc{S}_{Q}$. The scale invariance of the cross section is easily checked through the result, 
\be 
\gamma_h + \gamma_{hs} + 2(\gamma_c+\gamma_{cs}) = 0. 
\ee
In Eq.~\eqref{g1} and \eqref{g2}, $\Gamma_C$ is the cusp anomalous dimension~\cite{Korchemsky:1987wg,Korchemskaya:1992je}. We employed the first two terms in the expansion, $\Gamma_{C} = \sum_{k=0} \Gamma_{k}(\as/4\pi)^{k+1}$, where the two coefficients are given as 
\be
\Gamma_{0} = 4C_F,~~~\Gamma_{1} = 4C_F \Bigl[\bigl(\frac{67}{9}-\frac{\pi^2}{3}\bigr) C_A - \frac{10}{9} n_f\Bigr].
\ee 

Solving the RG equations in Eq.~\eqref{RGE}, we exponentiate large logarithms to NLL accuracy, and the result for the cross section is given by 
\be 
\label{resum}
\sigma_2(Q,\beta, R,m) = \exp\Bigl[\mc{M}(\mu_h,\mu_{hs},\mu_c,\mu_{cs})\Bigr] H(\mu_h) S_{hs}(\mu_{hs}) \bigl[\mc{J}_Q(\mu_c) 
\mc{S}_Q(\mu_{cs})\bigr]^2, 
\ee
where the factorization scale dependence in each factorized function in Eq.~\eqref{hqdijetf} has been exactly cancelled. 
Here we set the default scales for the evolutioned functions as $\{\mu_h^0,\mu_{hs}^0,\mu_{c}^0,\mu_{cs}^0\} = 
\{Q,2\beta Q, B,2\beta B\}$, where $B = \sqrt{(QR/2)^2+m^2}$.
The exponentiation factor in Eq.~\eqref{resum} is 
\bea 
\mc{M}(\mu_h,\mu_{hs},\mu_c,\mu_{cs}) &=& 4S_{\Gamma} (\mu_h,\mu_{hs})-4S_{\Gamma} (\mu_c,\mu_{cs}) + 2\ln\frac{\mu_h^2}{Q^2} a_{\Gamma} (\mu_h,\mu_{hs}) - 2\ln\frac{\mu_c^2}{B^2} a_{\Gamma} (\mu_c,\mu_{cs}) \nnb \\
\label{expfac}
&& \! \! \! \! \! \!
+4\ln2\beta~ a_{\Gamma} (\mu_{hs},\mu_{cs}) - \frac{2C_F}{\beta_0} \Bigl(\frac{3+b}{1+b} \ln\frac{\as(\mu_h)}{\as(\mu_c)}+\frac{2b}{1+b} \ln\frac{\as(\mu_h)}{\as(\mu_{cs})}\Bigr).
\eea 
Here $S_{\Gamma}$ and $a_{\Gamma}$ are defined as 
\be
S_{\Gamma} (\mu_1,\mu_2) = \int^{\alpha_1}_{\alpha_2} \frac{d\as}{b(\as)} \Gamma_{C}(\as) \int^{\as}_{\alpha_1}
\frac{d\as'}{b(\as')},~~~a_{\Gamma}(\mu_1,\mu_2) = \int^{\alpha_1}_{\alpha_2} \frac{d\as}{b(\as)} \Gamma_C(\as),
\ee
where $\alpha_{1,2} \equiv \as (\mu_{1,2})$, and $b(\as)=d\as/d\ln\mu$ is QCD beta function to be expanded as 
$b(\as)=-2\as\sum_{k=0}\beta_k (\as/4\pi)^{k+1}$.

The exponentiation of Eq.~\eqref{expfac} is not sufficient for the full resummation at NLL accuracy since it does not include large nonglobal logarithms~\cite{Dasgupta:2001sh,Banfi:2002hw}, which start to appear at order $\as^2$. In our case collinear gluon radiations from the heavy quark have a limited phase space bounded by $R$. Then the decoupled csoft gluons from the collinear gluon and the heavy quark can give rise to the nonglobal logarithms at the higher orders than order $\as$. Resummation of the nonglobal logarithms involved with a heavy quark is beyond the scope of this paper.\footnote{\baselineskip 3.0 ex Very recent study of nonglobal logarithm resummation related to top pair production~\cite{Balsiger:2020ogy} would be helpful for the future analysis. }

\begin{figure}[t]
\begin{center}
\includegraphics[height=10cm]{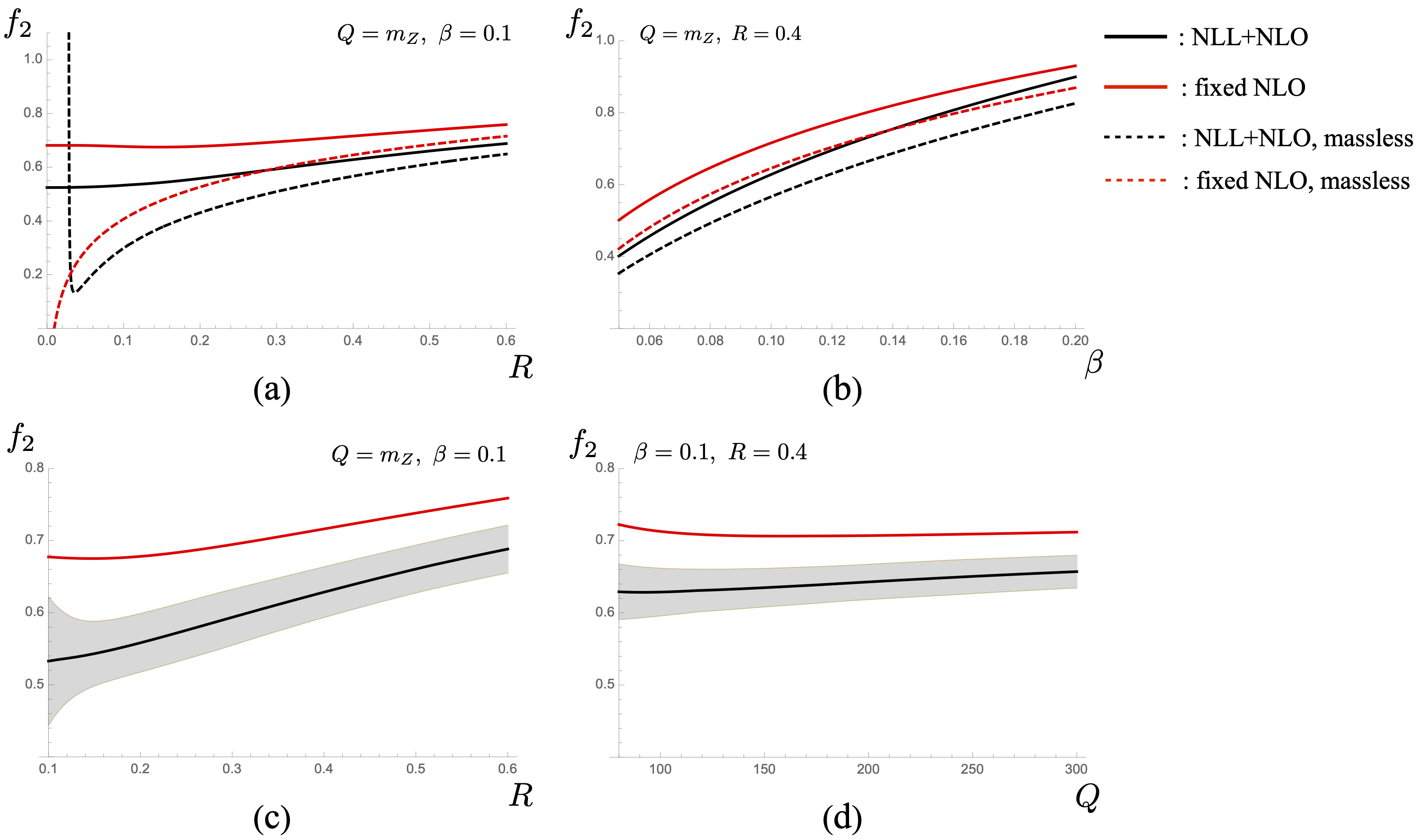}
\end{center}
\vspace{-0.8cm}
\caption{\label{fig2}\baselineskip 3.5ex
The $b$-dijet ratio with variations of $R$, $\beta$, and $Q$. Here black solid line denotes the resummed result with inclusion of the fixed NLO contribution, and the red solid line is the fixed NLO result without resummation. The dashed lines are the results in the massless limit. }
\end{figure}

For numerical implementation, we have considered the dijet ratio of $e^+e^-$-annihilation, i.e., $f_2 = \sigma_2/\sigma_{\mr{tot}}$. Here $\sigma_{\mr{tot}}$ is the inclusive cross section for the heavy quark pair production, and it is given as $\sigma_{\mr{tot}} = \sigma_0 (1+\as/\pi)$ to NLO in $\as$. 
In Fig.~\ref{fig2} the resummed results for the $b$ dijet ratio have been illustrated and compared with the fixed order results at NLO. The resummed result has significant suppression.   

In Fig.~\ref{fig2}~(a) and (b), the $b$ quark mass effects have been estimated by comparing with the results in the massless limit. The inclusion of the heavy quark mass enhances the results sizably. For example, the $b$ dijet ratio with $\beta = 0.1$ at $Z$ pole increases by $11-16~\%$ due to the quark mass under variation of $R\in[0.2,0.4]$. If we consider the charm dijet ratio in the same situation, the charm quark mass effect enhances it by $2-7~\%$. 

In Fig.~\ref{fig2}~(c) and (d), scale variations of the resummed result have been estimated. When we obtain the errors, we independently vary the scales, $\mu_i~(i=h,hs,c,cs)$, from $\mu_i^0/2$ to $2\mu_i^0$. The scale uncertainty from the four scale variations is rather large. In spite of this, we still observe meaningful deviations from the fixed order results. 
If we obtain the dijet ratio to higher order accuracy in the resummation,  
the uncertainty should be significantly reduced. This will be the focus of future work.   

\section{Exclusive heavy quark pair production}
\label{sec4}

If we look into Fig.~\ref{fig2} (a), we see that the heavy quark dijet cross section can be safely extended to the limit $R = 0$. Unlike the massless case, a collinear divergence does not arise in this limit due to the heavy quark mass. 
So, if we consider the exclusive heavy quark pair production, the IR safe cross section can be obtained from the dijet cross section taking the limit $R\to 0$. In this case, the jet veto with $\beta$ in the dijet cross section can be considered as an energy cut of soft hadrons. As a result, the cross section for the heavy quark pair can be regarded as the cross section for ``the hemisphere isolation of the heavy quark''.  

The factorization theorem for the exclusive heavy quark pair production can be immediately obtained from the result of the dijet cross section in Eq.~\eqref{hqdijetf}, and it leads to 
\be
\sigma_{Q\bar{Q}} (Q,\beta,m) = \sigma_0 H(Q,\mu) S_{hs}(2\beta Q,\mu) 
\Bigr[\mc{C}_m^2 (m,\mu) \mc{S}_{m}^2 (2\beta m,\mu) \Bigr].
\label{hqpf} 
\ee
Here, taking the limit of $\mc{J}_{Q}$ and $\mc{S}_{Q}$ as $R\to 0$, we obtain the collinear function $\mc{C}_m$ and the csoft function $\mc{S}_m$ respectively. The NLO results are given by
\bea
\label{cmnlo}
\mc{C}_m (m,\mu) &=& 1+ \frac{\as C_F}{2\pi} \Bigl[\frac{1}{2} \ln \frac{\mu^2}{m^2} +\frac{1}{2} \ln^2 \frac{\mu^2}{m^2} +2 +\frac{\pi^2}{12} \Bigr],  \\
\label{smnlo}
\mc{S}_m (2\beta m,\mu) &=& 1 + \frac{\as C_F}{2\pi} \Bigl[ \ln \frac{\mu^2}{4\beta^2m^2} -\frac{1}{2} \ln^2 \frac{\mu^2}{4\beta^2 m^2} -\frac{\pi^2}{12} \Bigr]. 
\eea 
Here $\mc{C}_m$ is the matching coefficient onto bHQET and the result of integrating out virtual collinear interactions of the heavy quark~\cite{Neubert:2007je,Fleming:2007xt,Fickinger:2016rfd}.  

\begin{figure}[t]
\begin{center}
\includegraphics[height=4.5cm]{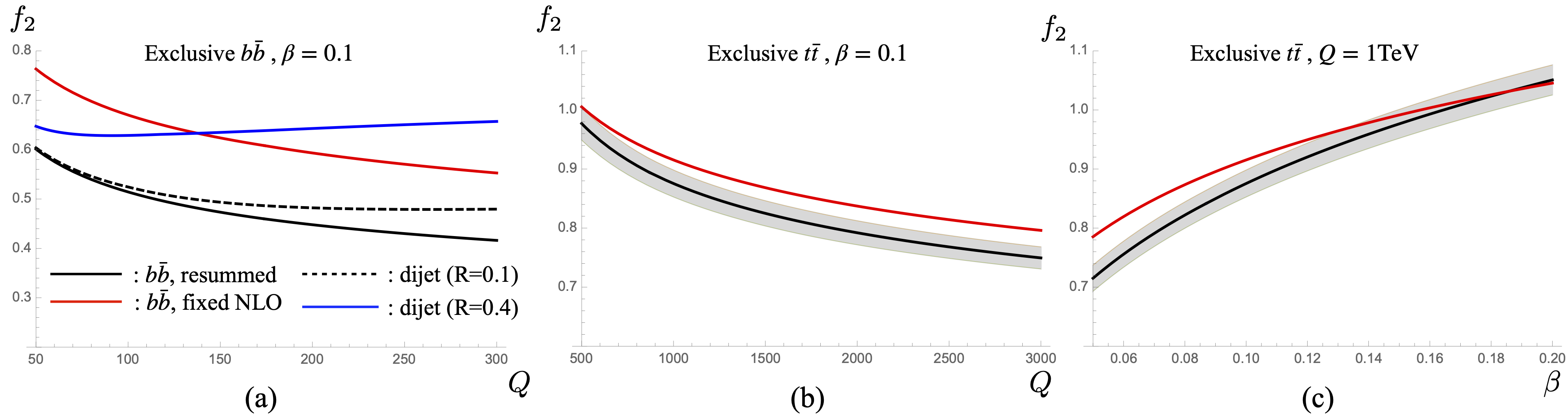}
\end{center}
\vspace{-0.8cm}
\caption{\label{fig3}\baselineskip 3.5ex
The exclusive heavy quark pair production fractions over the inclusive production. Red lines denote the non-resummed results. (a): $b$ quark pair production rates, and the dijet ratios with $R=0.1$ and $R=0.4$. (b-c): the fractions for the top pair productions with variations of $Q$ and $\beta$. Here all the resummed results are obtained at NLL accuracy and include the fixed NLO results.
}
\end{figure}

Using the factorization theorem in Eq.~\eqref{hqpf}, we resum the large logarithms of $Q/m$ and $\beta$ to NLL accuracy. 
The result is free from nonglobal logarithms since $R=0$. In Fig.~\ref{fig3}, we show the rate of the exclusive heavy quark pair production over total cross section for $Q\bar{Q}X$, which is defined as $f_2^{(Q)} = \sigma_{Q\bar{Q}}/\sigma_{\mr{tot}}$.
Like the dijet case, the exclusive cross sections are suppressed due to the resummation of large logarithms. 

In Fig.~\ref{fig3} (a), the resummed results for the $b$ quark pair production have been illustrated in the range $Q\in [50,300]~\mr{GeV}$. 
Compared with the resummed $b$-jet rate with $R=0.4$, the suppression of the $b$ quark pair production becomes larger as $Q$ increases. 
%The resummed rate for the charm is extremely small and the $b$ quark rate is less than 20$\%$ in most range. 
This is not surprising if we consider the dead cone effect~\cite{Dokshitzer:1991fc,Dokshitzer:1991fd,Ellis:1991qj}.  
As the heavy quark mass impact becomes smaller, the probability of collinear gluon radiations from the heavy quark become higher. In case of the exclusive production, in principle no collinear gluon radiation is allowed. Hence the rate for the $b$ quark pair production should be suppressed as the energy is large. 

The exclusive production for the $b$ quark is not realistic and the prediction here can be spoiled by nonperturbative interactions such as hadronization effects. Instead, for example, we may consider the dijet ratio with $R=0.1$ as shown in Fig.~\ref{fig3} (a) (the dotted lines). In this case the resummed results are more reliable, but still give small fractions, $f_2 \sim 0.5$. 
An interesting point is that the resummed results for both the cases describe very similar situation with a leading process at parton level, i.e., only $Q\bar{Q}$ in the final state. Here the fraction $f_2$ can be also considered as the ratio over LO cross section, $\sigma_0$, since NLO corrections to the total cross section are quite small. The smallness of $f_2$ implies that we cannot adhere to the view that the parton at leading process can be identified with a sharp jet.     

In Fig.~\ref{fig3} (b) and (c), the resummed results for the exclusive top pair production have been illustrated. Here the error bands have been estimated in the same way as the case of the dijet ratio. The top fraction is over $70\%$ in wide range of $Q$. This indicates that the top does not radiate many collinear gluons due to the large top quark mass. So, even with the extreme isolation of the top quark, we can expect rather a large cross section. 

\section{Conclusion}
\label{sec5}

We have considered the exclusive heavy quark dijet cross section. The factorization theorem is similar to the cross section for the massless case. But, the jet sector is modified to have the quark mass, and the factorization into the collinear and the csoft part can be systematically performed through matching between $\mr{SCET_M}$ and bHQET. 

Using the factorization theorem we obtained the resummed result for the heavy quark dijet cross section to NLL accuracy and compared it with the result in the massless limit. As a consequence, the heavy quark dijet ratios  become quite suppressed by the resummation of large logarithms, and the heavy quark mass effects sizably enhances the results compared to the massless limit unless $E_JR$ is much larger than the quark mass.  

Since the heavy quark mass removes the collinear divergence, we can investigate the extreme limit of the dijet cross section as $R\to 0$. The resulting cross section has been also analyzed to NLL accuracy. Compared with the LO result, the cross section for the exclusive top pair production is not suppressed much due to the large top quark mass, while the $b$ quark production is severely suppressed especially when the energy becomes large. The suppression through resummation of large logarithms implies that the results of some exclusive processes cannot be approximated as the LO results in $\as$ at the parton level. It would be interesting to apply this idea to exclusive leptonic processes with a tight energy cut of soft photons.  
\acknowledgments

The author is grateful to Adam Leibovich for useful comments. 
This study was supported by the Research Program funded by Seoul National University of Science and Technology.

%%%%%%%%%%%%%%%%%%%%%%%%%%%%%%%%%%%%%%%%%%%%%%%%%%%%%%%%%%%%%%%%%%%%%%
%%%%%%%%%%%%%%%%%%%%%%%%%%%%% Bibliography %%%%%%%%%%%%%%%%%%%%%%%%%%%
%%%%%%%%%%%%%%%%%%%%%%%%%%%%%%%%%%%%%%%%%%%%%%%%%%%%%%%%%%%%%%%%%%%%%%

%%%%%%%%%%%%%%%%%%%%%%%%%%%%%%%%%%%%%%%%%%%%%%%%%%%%%%%%%%%%%%%%%%%%%%


\begin{thebibliography}{99}
%\normalsize



%\cite{Dasgupta:2014yra}
\bibitem{Dasgupta:2014yra} 
  M.~Dasgupta, F.~Dreyer, G.~P.~Salam and G.~Soyez,
  %``Small-radius jets to all orders in QCD,''
  JHEP {\bf 1504}, 039 (2015)
  %doi:10.1007/JHEP04(2015)039
  [arXiv:1411.5182 [hep-ph]].


\bibitem{Kang:2016mcy} 
  Z.~B.~Kang, F.~Ringer and I.~Vitev,
  %``The semi-inclusive jet function in SCET and small radius resummation for inclusive jet production,''
  JHEP {\bf 1610}, 125 (2016)
%  doi:10.1007/JHEP10(2016)125
  [arXiv:1606.06732 [hep-ph]].

\bibitem{Dai:2016hzf} 
  L.~Dai, C.~Kim and A.~K.~Leibovich,
  %``Fragmentation of a Jet with Small Radius,''
  Phys.\ Rev.\ D {\bf 94}, no. 11, 114023 (2016)
%  doi:10.1103/PhysRevD.94.114023
  [arXiv:1606.07411 [hep-ph]].



%\cite{Bauer:2000ew}
\bibitem{Bauer:2000ew} 
  C.~W.~Bauer, S.~Fleming and M.~E.~Luke,
  %``Summing Sudakov logarithms in B ---> X(s gamma) in effective field theory,''
  Phys.\ Rev.\ D {\bf 63}, 014006 (2000)
 % doi:10.1103/PhysRevD.63.014006
  [hep-ph/0005275].


%\cite{Bauer:2000yr}
\bibitem{Bauer:2000yr} 
  C.~W.~Bauer, S.~Fleming, D.~Pirjol and I.~W.~Stewart,
  %``An Effective field theory for collinear and soft gluons: Heavy to light decays,''
  Phys.\ Rev.\ D {\bf 63}, 114020 (2001)
  %doi:10.1103/PhysRevD.63.114020
  [hep-ph/0011336].

\bibitem{Bauer:2001yt} 
  C.~W.~Bauer, D.~Pirjol and I.~W.~Stewart,
 % Soft collinear factorization in effective field theory,
  Phys.\ Rev.\ D {\bf 65}, 054022 (2002)
  [hep-ph/0109045].  

%\cite{Bauer:2002nz}
\bibitem{Bauer:2002nz} 
  C.~W.~Bauer, S.~Fleming, D.~Pirjol, I.~Z.~Rothstein and I.~W.~Stewart,
  %``Hard scattering factorization from effective field theory,''
  Phys.\ Rev.\ D {\bf 66}, 014017 (2002)
  %doi:10.1103/PhysRevD.66.014017
  [hep-ph/0202088].

\bibitem{Sterman:1977wj}
G.~F.~Sterman and S.~Weinberg,
%``Jets from Quantum Chromodynamics,''
Phys. Rev. Lett. \textbf{39}, 1436 (1977)
%doi:10.1103/PhysRevLett.39.1436


%\cite{Catani:1993hr}
\bibitem{Catani:1993hr} 
  S.~Catani, Y.~L.~Dokshitzer, M.~H.~Seymour and B.~R.~Webber,
  %``Longitudinally invariant $K_t$ clustering algorithms for hadron hadron collisions,''
  Nucl.\ Phys.\ B {\bf 406}, 187 (1993).
%  doi:10.1016/0550-3213(93)90166-M


%\cite{Ellis:1993tq}
\bibitem{Ellis:1993tq} 
  S.~D.~Ellis and D.~E.~Soper,
  %``Successive combination jet algorithm for hadron collisions,''
  Phys.\ Rev.\ D {\bf 48}, 3160 (1993)
%  doi:10.1103/PhysRevD.48.3160
  [hep-ph/9305266].

%\cite{Dokshitzer:1997in}
\bibitem{Dokshitzer:1997in} 
  Y.~L.~Dokshitzer, G.~D.~Leder, S.~Moretti and B.~R.~Webber,
  %``Better jet clustering algorithms,''
  JHEP {\bf 9708}, 001 (1997)
%  doi:10.1088/1126-6708/1997/08/001
  [hep-ph/9707323].
  
\bibitem{Cacciari:2008gp} 
  M.~Cacciari, G.~P.~Salam and G.~Soyez,
  %``The anti-$k_t$ jet clustering algorithm,''
  JHEP {\bf 0804}, 063 (2008)
%  doi:10.1088/1126-6708/2008/04/063
  [arXiv:0802.1189 [hep-ph]].  




\bibitem{Cheung:2009sg}  
  W.~M.~Y.~Cheung, M.~Luke and S.~Zuberi,
  %``Phase Space and Jet Definitions in SCET,''
  Phys.\ Rev.\ D {\bf 80}, 114021 (2009)
%  doi:10.1103/PhysRevD.80.114021
  [arXiv:0910.2479 [hep-ph]].

\bibitem{Ellis:2010rwa} 
  S.~D.~Ellis, C.~K.~Vermilion, J.~R.~Walsh, A.~Hornig and C.~Lee,
  %``Jet Shapes and Jet Algorithms in SCET,''
  JHEP {\bf 1011}, 101 (2010)
%  doi:10.1007/JHEP11(2010)101
  [arXiv:1001.0014 [hep-ph]].

\bibitem{Chay:2015ila} 
  J.~Chay, C.~Kim and I.~Kim,
  %``Factorization of the dijet cross section in electron-positron annihilation with jet algorithms,''
  Phys.\ Rev.\ D {\bf 92}, no. 3, 034012 (2015)
%  doi:10.1103/PhysRevD.92.034012
  [arXiv:1505.00121 [hep-ph]].

\bibitem{Becher:2015hka}
T.~Becher, M.~Neubert, L.~Rothen and D.~Y.~Shao,
%``Effective Field Theory for Jet Processes,''
Phys. Rev. Lett. \textbf{116} (2016) no.19, 192001
%doi:10.1103/PhysRevLett.116.192001
[arXiv:1508.06645 [hep-ph]].

\bibitem{Chien:2015cka}
Y.~T.~Chien, A.~Hornig and C.~Lee,
%``Soft-collinear mode for jet cross sections in soft collinear effective theory,''
Phys. Rev. D \textbf{93} (2016) no.1, 014033
%doi:10.1103/PhysRevD.93.014033
[arXiv:1509.04287 [hep-ph]].

%\cite{Leibovich:2003jd}
\bibitem{Leibovich:2003jd} 
  A.~K.~Leibovich, Z.~Ligeti and M.~B.~Wise,
  %``Comment on quark masses in SCET,''
  Phys.\ Lett.\ B {\bf 564}, 231 (2003)
%  doi:10.1016/S0370-2693(03)00565-3
  [hep-ph/0303099].

\bibitem{Rothstein:2003wh} 
  I.~Z.~Rothstein,
  %``Factorization, power corrections, and the pion form-factor,''
  Phys.\ Rev.\ D {\bf 70}, 054024 (2004)
%  doi:10.1103/PhysRevD.70.054024
  [hep-ph/0301240].

%\cite{Chay:2005ck}
\bibitem{Chay:2005ck} 
  J.~Chay, C.~Kim and A.~K.~Leibovich,
  %``Quark mass effects in the soft-collinear effective theory and anti-B ---> X(s gamma) in the endpoint region,''
  Phys.\ Rev.\ D {\bf 72}, 014010 (2005)
%  doi:10.1103/PhysRevD.72.014010
  [hep-ph/0505030].

\bibitem{DKL} 
  L.~Dai, C.~Kim and A.~K.~Leibovich, in preparation. 

%\cite{Dai:2018ywt}
\bibitem{Dai:2018ywt} 
  L.~Dai, C.~Kim and A.~K.~Leibovich,
  %``Heavy Quark Jet Fragmentation,''
  JHEP {\bf 1809}, 109 (2018)
%  doi:10.1007/JHEP09(2018)109
  [arXiv:1805.06014 [hep-ph]].   


%\cite{Korchemsky:1987wg}
\bibitem{Korchemsky:1987wg}
G.~Korchemsky and A.~Radyushkin,
%``Renormalization of the Wilson Loops Beyond the Leading Order,''
Nucl. Phys. B \textbf{283} (1987), 342-364
%doi:10.1016/0550-3213(87)90277-X

%\cite{Korchemskaya:1992je}
\bibitem{Korchemskaya:1992je}
I.~Korchemskaya and G.~Korchemsky,
%``On lightlike Wilson loops,''
Phys. Lett. B \textbf{287} (1992), 169-175
%doi:10.1016/0370-2693(92)91895-G


\bibitem{Manohar:2006nz} 
  A.~V.~Manohar and I.~W.~Stewart,
  %``The Zero-Bin and Mode Factorization in Quantum Field Theory,''
  Phys.\ Rev.\ D {\bf 76}, 074002 (2007)
%  doi:10.1103/PhysRevD.76.074002
  [hep-ph/0605001].

\bibitem{Dasgupta:2001sh} 
  M.~Dasgupta and G.~P.~Salam,
  %``Resummation of nonglobal QCD observables,''
  Phys.\ Lett.\ B {\bf 512}, 323 (2001)
%  doi:10.1016/S0370-2693(01)00725-0
  [hep-ph/0104277].

\bibitem{Banfi:2002hw} 
  A.~Banfi, G.~Marchesini and G.~Smye,
  %``Away from jet energy flow,''
  JHEP {\bf 0208}, 006 (2002)
%  doi:10.1088/1126-6708/2002/08/006
  [hep-ph/0206076].

\bibitem{Balsiger:2020ogy}
M.~Balsiger, T.~Becher and A.~Ferroglia,
%``Resummation of non-global logarithms in cross sections with massive particles,''
[arXiv:2006.00014 [hep-ph]].


\bibitem{Neubert:2007je}
M.~Neubert,
%``Factorization analysis for the fragmentation functions of hadrons containing a heavy quark,''
[arXiv:0706.2136 [hep-ph]].

\bibitem{Fleming:2007xt}
S.~Fleming, A.~H.~Hoang, S.~Mantry and I.~W.~Stewart,
%``Top Jets in the Peak Region: Factorization Analysis with NLL Resummation,''
Phys. Rev. D \textbf{77}, 114003 (2008)
%doi:10.1103/PhysRevD.77.114003
[arXiv:0711.2079 [hep-ph]].

\bibitem{Fickinger:2016rfd}
M.~Fickinger, S.~Fleming, C.~Kim and E.~Mereghetti,
%``Effective field theory approach to heavy quark fragmentation,''
JHEP \textbf{11} (2016), 095
%doi:10.1007/JHEP11(2016)095
[arXiv:1606.07737 [hep-ph]].


\bibitem{Dokshitzer:1991fc}
Y.~L.~Dokshitzer, V.~A.~Khoze and S.~I.~Troian,
%``Particle spectra in light and heavy quark jets,''
J. Phys. G \textbf{17} (1991), 1481-1492
%doi:10.1088/0954-3899/17/10/003

\bibitem{Dokshitzer:1991fd}
Y.~L.~Dokshitzer, V.~A.~Khoze and S.~I.~Troian,
%``On specific QCD properties of heavy quark fragmentation ('dead cone'),''
J. Phys. G \textbf{17} (1991), 1602-1604
%doi:10.1088/0954-3899/17/10/023

\bibitem{Ellis:1991qj}
R.~K.~Ellis, W.~J.~Stirling and B.~R.~Webber,
%``QCD and collider physics,''
Camb. Monogr. Part. Phys. Nucl. Phys. Cosmol. \textbf{8}, 1-435 (1996)




\end{thebibliography}
\end{document}